# A general sample size framework for developing or updating a clinical prediction model

Richard D Riley[1,2], Rebecca Whittle,[1,2] Mohsen Sadatsafavi,[3] Glen P. Martin,[4] Alexander Pate,[4] Gary S. Collins[5], Joie Ensor[1,2]

**Contact details:**

* corresponding author:

Professor of Biostatistics; e- mail: r.d.riley@bham.ac.uk

[1] Department of Applied Health Sciences, School of Health Sciences, College of Medicine and Health, University of Birmingham, Birmingham, UK

[2] National Institute for Health and Care Research (NIHR) Birmingham Biomedical Research Centre, UK.

[3] Respiratory Evaluation Sciences Program, Faculty of Pharmaceutical Sciences, The University of British Columbia, Vancouver, Canada

[4] Division of Informatics, Imaging and Data Science, Faculty of Biology, Medicine and Health, University of Manchester, Manchester Academic Health Science Centre, Manchester, UK

[5] Centre for Statistics in Medicine, Nuffield Department of Orthopaedics, Rheumatology and Musculoskeletal Sciences, University of Oxford, Oxford, OX3 7LD, UK

**Funding**: This paper presents independent research supported by an EPSRC grant for 'Artificial intelligence innovation to accelerate health research' (number: EP/Y018516/1), by an MRC-NIHR Better Methods Better Research grant (MR/Z503873/1) and by the National Institute for Health and Care Research (NIHR) Birmingham Biomedical Research Centre at the University Hospitals Birmingham NHS Foundation Trust and the University of Birmingham. GSC is supported by Cancer Research UK (programme grant: C49297/A27294). PD is supported by Cancer Research UK (project grant: PRCPJT-Nov21\100021). LK is supported by an NIHR Doctoral Fellowship (NIHR303331) and by a core funding grant awarded to the Cancer Research UK Clinical Trials Unit by Cancer Research UK (CTUQQR-Dec22/100006). RDR and GSC are National Institute for Health and Care Research (NIHR) Senior Investigators.  The views expressed are those of the authors and not necessarily those of the NHS, the NIHR or the Department of Health and Social Care.

**Competing Interests**: RDR receives royalties for sales of his textbooks "Prognosis Research in Healthcare" and "Individual Participant Data Meta-analysis".

**Word count:** 7400

# Abstract

Background

Existing sample size calculations for developing clinical prediction models rely on standard (unpenalised) regression-based formula. A general approach is needed for developing or updating a CPM using any statistical or machine learning method.

Objectives

To propose a general framework for sample size calculations applicable to any model development approach, based on drawing samples from anticipated sampling/posterior distributions and targeting minimal degradation in predictive performance compared to reference model.

Methods

Researchers must provide candidate predictors, their reference model (e.g., a regression equation matching outcome incidence, predictor weights and c-statistic of previous models), and a (synthetic) dataset reflecting the joint distribution of candidate predictors in the target population. Then, a fully simulation-based approach allows the impact of a chosen development sample size and modelling strategy to be examined. This generates thousands of models and, by applying each to the entire target population, leads to sampling/posterior distributions of individual predictions and model performance (degradation) metrics, to inform required sample sizes. To improve computational speed for penalised regression, we also propose a one-sample Bayesian analysis that combines shrinkage priors with a likelihood decomposed into sample size and Fisher's unit information.

Results

The framework is illustrated when developing pre-eclampsia prediction models using logistic regression (unpenalised, uniform shrinkage, lasso or ridge) and random forests. It encompasses existing sample size calculation criteria whilst also providing model assurance probabilities, instability metrics, and degradation statistics about calibration, discrimination, clinical utility, prediction error and fairness. Crucially, the required sample size depends on the users' key estimands of interest and planned model development or updating approach.

Conclusions

The new, flexible framework generalises existing sample size proposals for model development or updating by utilising anticipated sampling/posterior distributions conditional on a chosen sample size and development strategy. This informs the sample size required to target appropriate model performance, stability, clinical utility and fairness.

# 1. Introduction

Studies that develop a clinical prediction model (CPM) use a sample of data, representative of a target population (e.g., women diagnosed with breast cancer), to produce a model for estimating an individual's outcome risk (e.g., 5-year risk of breast cancer recurrence) conditional on their values of multiple predictors (e.g., age, stage of disease). An example is the ISARIC model,[1] for use in hospitalised adults with suspected or confirmed COVID-19 to estimate their risk of in-hospital clinical deterioration based on 11 predictors measured at hospital admission.

Many papers have proposed sample size calculations for studies developing a CPM. Initially these stemmed from rules-of-thumb such as having 10 events per predictor,[2] but in recent years more tailored equations and approaches have emerged for determining the sample size required to target precise predictions, low overfitting and instability, small mean square prediction error, and a small loss of clinical utility.[3-9] However, this work is typically based on standard (unpenalised) regression-based approaches, and so in this article we propose a more general framework for developing or updating a CPM using any statistical or machine learning method. Our approach aims to anticipate the sampling (posterior) distributions of individual-level predictions for a particular sample size, in the context of a particular model development strategy and included set of candidate predictors. Then, by repeated sampling from these distributions, the impact of a particular sample size can be examined in terms of model performance metrics. These may include measures of expected: (i) *model degradation*, which summarise the anticipated reduction in model performance (e.g., calibration, discrimination and clinical utility of estimated risks) compared to an assumed reference model (e.g., based on previous models in the field); (ii) *prediction instability*, for example as defined by the mean width of 95% uncertainty intervals around individual risk estimates; and (iii *model assurance*, defined by the probability that a single dataset will give a CPM that meets some pre-defined criteria (e.g., calibration slopes between 0.9 and 1.1; relative value of sample information for net benefit > 90%).

The article outline is as follows. Section 2 describes the overall premise and requirements for sample size calculations for prediction model development. Section 3 outlines the new framework, which follows a fully simulation-based approach to examine model degradation, instability and assurance relative to a reference model. To substantially improve computational time, Section 4 proposes an approximation for penalised regression based on a Bayesian one-sample approach using a decomposition of Fisher's information matrix. Section 5 considers extension to a fully Bayesian approach to allow for uncertainty in the reference model parameters. Section 6 concludes with some discussion. To illustrate the approaches, throughout we consider the sample size required for development or updating models for predicting adverse outcomes in pre-eclampsia.

## 2. Premise of sample size calculations for CPM development studies

In this section we outline the premise of sample size calculations for CPM development studies, to lay foundations for Sections 3 and 4. We focus on the prediction of binary outcomes in this article, but the general issues also apply to other outcomes (e.g., continuous or multinomial). We refer to model *development*, but stress the framework also covers model *updating* situations.

**(a) The calculation should address the key estimand of the study**

When developing a CPM for a binary outcome conditional on a set of candidate predictors, the estimand of interest is the risk for each individual in the target population. Therefore, for the sample size calculation, the focus is on how sample size impacts a CPM's sampling distribution (or posterior distribution in a Bayesian context) of risk for each individual from the target population. A sampling/posterior distribution for individual risk reflects the full information from the CPM (and predictors therein), for example in terms of the mean or median value of risk (best estimate of risk) and the uncertainty (e.g. 95% range) of plausible values.

In this article, we refer to uncertainty distributions (i.e., sampling and posterior distributions) broadly as posterior distributions, and define the $i^{\text{th}}$ individual's posterior distribution by $f(p_i \mid \boldsymbol{\beta}, \boldsymbol{x}_i)$ where $\boldsymbol{\beta}$ refers to the parameters of some true (but unknown) CPM, $\boldsymbol{x}_i$ denotes the individual's values of the predictors in the model and $p_i$ denotes the individual's true (underlying) risk conditional on $\boldsymbol{\beta}$ and $\boldsymbol{x}_i$.

**(b) The calculation should have target values for statistical performance and assurance**

Small sample size reduces model performance in terms of overall fit, calibration, discrimination and clinical utility, and leads to large uncertainty in individual predictions (reflecting model instability).[10] In other words, smaller sample sizes increase epistemic uncertainty (model-based error) and degrade a model's performance in the target population compared to if we knew the model perfectly (no model-based error). Therefore, in the target population of interest, and for the specific set of predictors considered in the model development, we can define,

model degradation = (developed model performance) – (true model performance)

where the true model performance is that which would be obtained if we had perfect information (i.e., an infinite sample size) about the predictor-outcome relationships.[5] Related concepts are the loss function and value of perfect information.[11] The sample size for a model development study should target an acceptable level (as defined by the researchers in their context of interest) of *model degradation*, in terms of the performance measures defined in Figure 1.

In particular, researchers may identify the (minimum) sample size needed to target a (maximum) acceptable value for the *expected* deterioration in key performance measures of interest, such as the expected reduction in the c-statistic (a measure of discrimination) or net benefit (a measure of clinical utility),[12] or the expected calibration slope (ideal value of 1). Similarly, the sample size may target expected levels of uncertainty and predictor error, including the expected absolute difference between risk estimates from fitted and true models (i.e., the mean absolute prediction error (MAPE)), the expected uncertainty (e.g., mean width of 95% uncertainty intervals) around risk estimates, and the expected effective sample size for individual predictions.[13]

Rather than targeting *expected* values (i.e., what is anticipated on average, such as a calibration slope of 1 or a reduction in c-statistic of 0.02), an alternative is to target *assurance* defined by a large probability of meeting desired performance criteria. Assurance probabilities are well-known for trial design,[14] but here we propose them for model development studies. For example, researchers might consider sample sizes to ensure a probability $\geq$ 0.9 that a developed CPM will have a calibration slope between 0.9 to 1.1 in the target population; a probability $\geq$ 0.9 that the degradation in c-statistic will be < 0.025; or a probability of $\geq$ 0.9 that the degradation in net benefit will be $\leq$ 10%.

Those designing a CPM development study must prioritise which performance measures (e.g., calibration, discrimination and net benefit) and target metrics (e.g., expected values, assurance probabilities) are most relevant for their setting and purpose; this requires a multidisciplinary consultation including clinicians and patients. For example, if accurate predictions are required at the individual level (e.g., adverse outcome prediction in women with pre-eclampsia), then a strong emphasis may be placed on targeting strong calibration and small uncertainty interval widths around individual risk estimates. Conversely, if the focus is population-level improvements (e.g., reduced CVD incidence in the population), then assurance might focus more on population-level measures such as net benefit. Crucially, any measure of interest can be derived directly from posterior (sampling) distributions of individuals' predictions; hence, our framework later focuses on deriving and utilising anticipated posterior distributions for a given sample size.

### (c) The calculation should be compatible with the planned analysis approach

Posterior distributions for individual-level predictions will depend on the modelling approach used to develop the CPM, and so any *general* sample size calculation needs to be flexible enough to allow *any* modelling approach to be included (e.g., regression, random forests, neural networks). In Section 3, we propose a simulation-based approach for this. However, for some modelling options like regression, the anticipated posterior distributions can be closely approximated using statistical theory, to give analytic solutions that allow a faster (computationally less-demanding) approach. In Section 4 we demonstrate this approach for (Bayesian penalised) regression approaches.

*Figure 1: Predictive performance measures of interest for examining expected model degradation due to using a particular sample size for CPM development*

> Compared to the predictions and performance of a 'true model' (i.e., the model that would be obtained had we an infinite sample size to apply our chosen development approach on the same set of candidate predictors), a CPM development study may desire a sample size that targets acceptable levels for the following measures:
>
> - **reduction in the percentage of explained outcome variation in the target population**; for example, as measured by $R^2$ (percentage of outcome variance explained by the model), including generalised $R^2$ measures for binary outcomes such as the Cox-Snell or Nagelkerke $R^2$.
>
> - **reduction in discrimination performance in the target population;** discrimination measures the separation of a model's estimated risks between those with and without the event, typically measured by the c-statistic (equivalent to AUROC for binary outcomes)
>
> - **mean square or absolute difference in estimated and 'true' model predictions in the target population**; this measures the model-based error (epistemic uncertainty), as the difference between an individual's 'true model' prediction $p_i$ (had the model been derived on perfect information) and estimated prediction $\hat{p}_i$ (from the CPM derived with a particular sample size). For example, given a population of size ($n_T$) in the target population, we can define these measures as:
>
>   - the root mean squared prediction error (RMSPE) = $\sqrt{\frac{\sum_{i=1}^{n_T}(\hat{p}_i - p_i)^2}{n_T}}$
>
>   - the mean absolute prediction error (MAPE) = $\frac{\sum_{i=1}^{n_T}|\hat{p}_i - p_i|}{n_T}$
>
> - **(mis)calibration of predictions in the target population;** calibration measures the agreement between observed and estimated risks, and researchers should quantify the magnitude of any miscalibration. In particular, fitting a logistic regression model in the target population comparing the observed outcome event ($Y_i = 0$ (no event) or 1 (event)) to the CPM's estimated predictions ($\hat{p}_i$) on the logit scale, then
>
>   $$Y_i \sim \text{Bernoulli}(p_i) \qquad \text{logit}(p_i) = \alpha + \beta \, \text{logit}(\hat{p}_i)$$
>
>   and $\beta$ is the calibration slope (ideal value of 1). Moreover, calibration can be assessed graphically using a calibration plot that compares observed and predicted values via a flexible (non-linear) calibration curve fitted using a smoother or splines.
>
> - **uncertainty in the model's predictions**; for example, as measured by the width of 95% intervals around the calibration curve; the width of 95% (credible) intervals for each individual's risk from their posterior distribution, or the effective sample size for each individual's prediction.
>
> **Targeting expected values in the population:** For each measure, we might consider the sample size needed to target a particular *expected value*. That is, to seek the sample size so that the CPM is expected to have (i.e., on average) a chosen value (e.g., target a particular mean reduction in c-statistic, mean calibration slope, mean 95% interval width for individual risk or calibration curve) when applied in the target population.
>
> **Targeting assurance:** Alongside the expected value, the variability (e.g., standard deviation of the reduction in c-statistic, instability of the calibration curve) of values may also be of interest because smaller variability gives more assurance that a single dataset of that size will lead to a reliable CPM. This assurance can be formally quantified as a probability that a particular performance is achieved in the target population. For example, we might consider the sample size needed to develop a CPM that meets the following assurance criteria in the target population:
>
> - a probability of >90% that the calibration slope is between 0.9 and 1.1
> - a probability of >90% that the MAPE or RMSPE is < 0.05
> - a probability of >90% that the reduction in c-statistic is <0.025

Figure 2: Classification performance measures of interest for examining expected model degradation due to using a particular sample size for CPM development

> In situations where a risk threshold is of interest for clinical decision making (e.g., to decide when to initiate treatment, refer for biopsy, etc.), then other measures linked to classification are also relevant for quantifying model degradation, including:
>
> - **expected probability of misclassification (P(misclassification))**; that is, the proportion of an individual's posterior distribution that falls on the opposite side of the risk threshold compared to their 'true' prediction.[15]
>
> - **expected reduction of clinical utility in the target population**, for example as measured by the reduction in net benefit. This is outlined by Sadatsafavi et al.,[5] who highlight that the 'true' model (i.e., knowing each individual's true risk, $p_i$, based on the chosen set of predictors) provides a maximum net benefit ($NB_{max}$) that can be achieved (also known as 'expected value of perfect information (EPVI)'). Then, for a model estimated on a particular sample size, the net benefit of that model ($NB_{model}$) will be lower than ($NB_{max}$) due to misclassification compared to their 'true' prediction (see previous bullet point); when the probability of misclassification increases due to larger sampling error (lower sample size), the reduction in net benefit will increase. The mean difference in the fitted model's net benefit and $NB_{max}$ in the target population, provides the model's expected degradation in clinical utility for a particular sample size. Also known as the model's 'expected value of sample information' (EVSI). Alongside this measure of absolute degradation, the *relative* amount of degradation can be summarised as $100 \times (NB_{model}/NB_{max})$ and taking the mean gives the 'expected relative value of sample information' (ERVSI) for the model's net benefit.
>
> The value of sampling information might also account for other strategies such as 'treat none' and 'treat all' (or other models), alongside the model of interest. In particular, based on observed results in the development sample, the strategy with the highest observed net benefit would be the one selected as the 'winner'. This can vary across samples by chance, and so the true value of sample information is how the winner's net benefit ($NB_{winner}$) degrades compared to that of the true model.
>
> **Targeting assurance:** Alongside the *expected* value of sample information, the *variability* may also be of interest because smaller variability gives more assurance that a single dataset of that size will be closer to the expected value. This assurance can be formally quantified as a targeted probability; for example:
>
> - A probability of 90% that a randomly selected individual has P(misclassification) < 5%
> - A probability of 90% that $RVSI_{model} \geq 90\%$
> - A probability of 90% that $RVSI_{winner} \geq 90\%$

## 3. A general framework for sample size calculations when developing a CPM

We now introduce a general framework for calculating and examining the sample size required for CPM development studies. The basis stems from Section 2, namely: (i) anticipating the posterior distribution of individual-level predictions that would arise when developing a CPM using a chosen modelling approach with a particular sample size; and subsequently (ii) obtaining and summarising the posterior distributions of this CPM's predictions and performance in the target population (e.g., using the measures of performance and degradation described in Figure 1 and Figure 2).

### 3.1 Step-by-step guide to a simulation-based approach to examining sample size

As there are a wide variety of model development approaches of interest for CPM researchers, and closed-form (parametric) approximations for posterior distributions are challenging to anticipate for some approaches (e.g., random forests or deep learning), our general approach uses simulation. Simulation is not a new idea for sample size calculations;[16] notably, Harrell's textbook illustrates how to use simulation for examining the minimum sample size required to target a particular MAPE,[17] assuming a true model with one continuous predictor. Below, we generalise this to include multiple predictors and provide ten steps for how to sample from posterior distributions to summarise the CPM's anticipated performance and degradation in the target population, compared to an assumed 'true model'. Here onwards, we refer to this 'true model' as a *reference model*, which is a working model (e.g., based on previous models in the field) specified by the user to implement our approach.

**PART A: Set-up phase to produce datasets for development and testing in the target population**

**Step (1) - specify the candidate predictors and any variables linked to fairness checks**

All candidate predictors that will be used to develop the CPM need to be specified. This should include *core predictors* that are well-established as important predictors of the outcome, and any exploratory candidate predictors (e.g., additional biomarkers). Also, variables linked to fairness checks should be specified, even if they are not candidate predictors (e.g., those representing protected characteristics and subgroups). Note that, if planning to use an existing dataset for CPM development, only the predictors and variables in that dataset can be chosen.

**Step (2) – specify the joint distribution of the predictors and variables from step (1)**

The joint distribution of predictors and variables defines the case-mix of the target population of interest. This can be based on that observed in an existing dataset or a pilot study in the same target population. Sometimes, the development dataset may exist but is unavailable. In this instance, the data holders could be asked to provide a synthetic dataset that mimics the joint distribution, for instance using a simulation-based approach that models conditional relationships,[18] such as via *synthpop* in R.[19] For example, Clinical Practice Research Datalink (CPRD) generates synthetic data to improve workflows (https://www.cprd.com/synthetic-data). Sometimes there will be no available data and no joint distribution information; in this case it may be necessary to assume conditionally independent distributions for the predictors. Previous work suggests this can be a reasonable approximation when focusing on some measures of model degradation such as expected calibration slope and MAPE.[6, 7] Different plausible assumptions of the joint distribution can be tried to explore the impact on sample size.

**Step (3) - specify a** reference model

To examine the impact of sample size on model degradation, a crucial step is to specify a reference model that expresses (a transformation of) the outcome as a function of assumed effects of the predictors from step (2). Like any sample size calculation, this requires drawing on previous evidence and judgement, for example from existing studies, previous models in the field and clinical expertise. Even if a 'black box' CPM development approach (e.g., random forests, neural networks) is planned, we recommend specifying the reference model as a regression equation, as it is pragmatic and interpretable. For example, a logistic regression model could be specified as,

$$y_i \sim \text{Bernoulli}(p_i)$$
$$\ln\left(\frac{p_i}{1-p_i}\right) = \mu_i = \alpha + \delta(\beta_1 x_{1i} + \beta_2 x_{2i} + \cdots + \beta_P x_{Pi})$$

Eq. (1)

with the user defining the $P$ parameter values to match anticipated relative predictor weights and model performance in the target population. The $\beta$ coefficients provide relative weights of predictors ($x$). In the absence of other information, and assuming all candidate predictors will add predictive information, each predictor might be standardised and all $\beta$ values set to 1. Then, if the user provides the anticipated c-statistic and overall risk, the iterative process of Austin[20] can be used to identify values of $\alpha$ and $\delta$ that match these (within a small margin of error) in the target population defined by step (2). Note that, to allow for added complexity that machine learning approaches aim to model, the regression model could include additional parameters with interactions, categorised continuous predictors, and non-linear relationships.

When users are planning to consider candidate predictors for which little or no evidence exists about their (added) predictive value, it is conservative to exclude them from the reference model. For example, if developing a cancer prognostic model with a set of well-known predictors (e.g., age and stage of disease) and 10 exploratory biomarkers, then we recommend excluding those 10 biomarkers from the reference model but still modelling them as noise variables in the steps that follow.

**Step (4) – generate a large dataset that represents the target population**

To form a large dataset representative of the target population, randomly sample candidate predictor values from the joint distribution (or the available synthetic dataset) specified in step (2) for a large number of hypothetical individuals. Then, use the reference model (from step (3)) to calculate the true risk (i.e., $p_i$) for each individual conditional on their predictor values. Randomly generate their observed outcome value (i.e. $y_i$ = 1 if event, $y_i$ = 0 if no event) from $y_i \sim \text{Bernoulli}(p_i)$. We suggest the dataset typically contains at least 100,000 observations, with thousands of outcome events. This allows the performance of the models to be estimated with very little error. However, the most appropriate dataset size will depend on the prevalence of the outcome while also considering the additional computational time of a larger dataset.

**Step (5) – generate a development dataset of size $n$ and other relevant samples (e.g., tuning datasets)**

Follow the same approach as in Step (4), to create a development dataset of $n$ individuals containing their (candidate) predictor and outcome values. If using an existing dataset for model development, then $n$ is fixed. Otherwise, when planning prospective data collection, this is the user's chosen sample size of interest, perhaps stemming from previous sample size recommendations.[4] Sometimes further samples are used toward model building, such as hold-out samples for tuning and recalibration; if so, these should also be randomly generated for a specified size. Users may also consider the approach to handling missing data here, but we do not consider this further in this article.

**PHASE B: Develop one CPM using the development dataset and quantify performance in the target population dataset**

**Step (6) – develop a CPM applying a chosen modelling approach to the development dataset**

Use the development dataset of size $n$ from step (5) to build a CPM using the planned model development approach, ensuring any user-specific choices are properly reflected (e.g., the approach to variable selection and hyperparameter tuning; how a model is recalibrated after testing in a hold-out sample; adjustment for optimism via bootstrapping; any sampling approaches to handle class imbalance, such as SMOTE and subsequent miscalibration adjustments[21]).

**Step (7) – apply the developed CPM to make predictions for each individual in the target population dataset**

Apply the CPM developed in step (6) to each individual in the target population dataset from step (4) to obtain a prediction (estimated value, $\hat{p}_i$) for their risk ($p_i$), conditional on their observed predictor values (i.e., $x_{1new}, x_{2new}, \ldots, x_{Pnew}$).

**Step (8) – quantify the CPM's predictive performance and degradation in the target population dataset, both overall and in any subgroups linked to fairness checks**

Calculate the CPM's predictive performance in the large target population dataset from step (4), and the degradation in performance compared to the reference model. The performance measures of interest should be chosen by the user, but generally we recommend considering discrimination, calibration, and net benefit (see Figure 1 and Figure 2). These can also be examined within specific subgroups of interest, such as those defined by ethnicity, as part of any fairness checks.

**PHASE C: Generate and summarise posterior distributions of CPM parameters, predictions, performance and degradation**

**Step (9) – repeat steps (5) to (8) many times and summarise the posterior distributions of predictions, predictive performance and model degradation in the target population dataset**

Repeating steps (5) to (7) involves repeatedly sampling a development dataset (of the chosen size, $n$) and building a separate CPM to each, using the same development approach each time. This leads to multiple CPMs, essentially obtained by repeatedly sampling from the joint posterior distribution of the model parameters. We recommend at least 1000 samples to ensure the Monte Carlo (MC) simulation error is relatively small, but ideally more if the computational time is acceptable. We refer to Morris et al. for more consideration of simulation size and MC error.[22]

Applying each of these (e.g., 1000) CPMs to each individual in the target population dataset generates multiple predictions (i.e. multiple $\hat{p}_i$ values) per individual, reflecting their posterior distribution (i.e. $p(p_i | \boldsymbol{\beta}, \boldsymbol{x}_{new})$). From this, uncertainty measures can be derived (such as a 95% interval for each individual's true $p_i$) and displayed (e.g., using a prediction instability plot comparing estimated versus true risks).[15]

Repeating step (8) many times also produces posterior distributions for measures of CPM performance and degradation in the target population. These can be summarised in terms of their mean (expected) value and uncertainty (e.g., mean and 95% interval for calibration slope; the probability that the degradation in net benefit will be $\leq$ 10%). They can also be displayed, for example, in an instability plot of calibration curves. Again, it is entirely the user's decision as to which measures to focus on.

**PHASE D (OPTIONAL): Examine other sample sizes and reference models**

**Step (10): Repeat steps (5) to (9) for alternative development sample sizes and assumptions**

Unless using an existing dataset with a fixed $n$ for model development, most users will want to examine a range of sample sizes, particularly when planning prospective data collection. In this instance, they will want to identify a (minimum) sample size that is expected to meet some target criteria of their choice (e.g., mean degradation in c-statistic < 0.02; mean degradation in net benefit < 10%; probability ≥ 90% that calibration slope ≥ 0.9 and ≤ 1.1; probability of 90% that degradation in net benefit ≤ 10%;). Similarly, they may want to repeat the process with different (yet still plausible) reference models and identify how the required sample size changes. Section 4 considers how to account for the uncertainty in the reference model, within step (3) itself.

### 3.2 Applied example: Prediction of adverse outcomes in pre-eclampsia

Now, we apply the full simulation-based approaches to estimate the sample size required to develop or update a prediction model for the risk of adverse outcomes (by hospital discharge) in pregnant women diagnosed with pre-eclampsia. Although previous models exist in this field, risk prediction in pre-eclampsia remains an area of active research. Here, we build on Thangaratinam et al. as a foundation,[23, 24] which reports a penalised regression model based on 9 predictors and an overall outcome risk of 0.68. The question is: what (minimum) sample size is needed to reliably build a new model with these nine predictors (and potentially others), that extends (updates) previous models in this field?

*3.2.1 Simulation set-up, model development approaches and performance measures*

We applied the 10-step process of Section 3.1 to implement the simulation-based sample size calculation, for each of the following development methods separately: an unpenalised logistic regression (no variable selection), an unpenalised logistic regression followed by a heuristic uniform shrinkage,[25] penalised logistic regression with either a ridge or lasso penalty in both frequentist (using 10-fold cross validation for tuning),[26, 27] Bayesian frameworks (for the chosen prior distributions, see Section 4.2), and a random forest with 100 trees and depth of either 3 (low) or 15 (high) for contrast. For brevity, a summary of our 10 steps is presented in Figure 3 and Stata code is available at https://github.com/Richard-D-Riley/code. The reference model was based on previous models in the field, and included 10 parameters for the 9 predictors, with a corresponding c-statistic of 0.76 and a net benefit of 0.41 at a risk threshold of 0.5 suggested by Thangaratinam et al. for decision making (e.g., transfer to tertiary units if risk ≥ 0.5).[24]

*Figure 3: The 10-step process of the fully simulation-based sample size calculation, as applied to a new study aiming to develop a prediction model for the risk of adverse outcomes (by discharge) in pregnant women with pre-eclampsia.*

**Step (1) - specify the candidate predictors and any variables linked to fairness checks**

Building on Thangaratinam et al.,[23, 24] we focus initially on nine predictors of: maternal age at pregnancy (years), gestational age at diagnosis (weeks), medical history (i.e., none, 1, or 2+ conditions), urine protein creatinine ratio (mg/mmol), serum urea (mmol/L), serum creatinine (µmol/L), systolic blood pressure (mmHg), parenteral anti-hypertensive therapy, and parenteral magnesium sulphate administered before or within 24h of diagnosis.

**Step (2) – specify the joint distribution of the predictors and variables from step (1)**

The original authors provided a large (~1.6 million) synthetic dataset, generated via *synthpop* in R,[19] that closely reflected their original dataset in terms of the joint distribution of predictors, with an overall outcome incidence (risk) of 0.68.

**Step (3) - specify a reference model**

Based on previous literature, we specified the following reference model with nine predictors (10 parameters):

$$\ln\left(\frac{p_i}{1-p_i}\right) = 14.8246 - (0.204 \times \text{age}) - (5.2265 \times \ln(\text{gestational age}))$$
$$- (0.3243 \times (1 \text{ previous condition})) - (0.6236 \times (\geq 2 \text{ previous conditions}))$$
$$+ (0.1665 \times \ln(\text{creatinine ratio})) + (0.4574 \times \ln(\text{urea})) - (0.0038 \times \text{creatinine})$$
$$+ (0.0232 \times \text{systolic blood pressure}) + (0.4552 \times \text{antihypertensive treatment})$$
$$+ (1.1425 \times \text{magnesium sulphate treatment})$$

In the target population dataset (see below) this reference model has a c-statistic of 0.76, a calibration slope of 1 and a net benefit of 0.41 at a risk threshold of 0.5 considered relevant for decisions.

**Step (4) – generate a large dataset that represents the target population**

We took a random sample of 100,000 individuals from the provided synthetic dataset and reserved these as our target population dataset. We randomly generated each individual's observed outcome value (i.e. $y_i$ = 1 if event, $y_i$ = 0 if no event) as $y_i \sim \text{Bernoulli}(p_i)$, with $p_i$ defined by the reference model of Step (3).

**Step (5) – generate a development dataset of size $n$ and other relevant samples (e.g., tuning datasets)**

We took a random sample of size $n$ from the ~1.5million individuals that remained in the synthetic dataset (i.e. those not moved to the target population dataset) and randomly generated their observed outcome value as $y_i \sim \text{Bernoulli}(p_i)$, where $p_i$ defined by the reference model of Step (3). We initially consider $n$ = 456, the minimum sample size estimated to be required to target a calibration slope of 0.9.

**Step (6) – develop a CPM applying a chosen modelling approach to the development dataset**

We consider various approaches for fitting the CPM: unpenalised logistic regression with and without subsequent uniform shrinkage; logistic regression with either a ridge or lasso penalty estimated in either a frequentist or Bayesian framework; and random forests. For the ridge and lasso approaches, 10-fold cross-validation was used to estimate the tuning parameter (lambda) when using a frequentist framework, and in the Bayesian framework the penalty was induced by the chosen priors shown in Section 4, with thinning and a burn-in of 5000 samples. When fitting all the models, we standardised predictors.

**Step (7) – apply the developed CPM to make predictions for each individual in the target population dataset**

For each CPM developed in step (6), we applied the fitted model to estimate risk for each of the *N* individuals.

**Step (8) – quantify the CPM's predictive performance and degradation in the target population dataset, both overall and in any subgroups linked to fairness checks**

We examine (degradation in) each model's performance in terms of discrimination (c-statistic), calibration (calibration slope and curve) and clinical utility (net benefit at a risk threshold of 0.5), as all these were deemed important and relevant for either patients, clinicians or methodologists.

**Step (9) – repeat steps (5) to (8) many and summarise the sampling (posterior) distributions of predictions, predictive performance and model degradation using the target population dataset**

We repeated steps (5) to (8) 999 times; then for each CPM approach, we used the 1000 sets of performance results to summarise (degradation) in discrimination, calibration and clinical utility. We also produced calibration, prediction and classification instability plots.[15]

**Step (10): Repeat steps (5) to (9) for alternative development sample sizes and assumptions**

We repeated the whole process considering a range of other sample sizes; and the impact of an additional 10 candidate continuous predictors (all standardised and independent) that are noise variables.

Our focus is on the anticipated (degradation in) CPM performance in terms of discrimination (c-statistic), calibration (calibration slope and curve) and clinical utility (net benefit at a risk threshold of 0.5); and the MAPE and 95% interval width around individual risk estimates. For brevity, we initially focus our evaluations on three sample sizes: $n$ = 456, 335 and 75. A sample size of $n$ = 456 (about 310 events and 31 events per candidate predictor parameter for regression models) is the *minimum* recommended by *pmsampsize*,[4, 28, 29] that targets a calibration slope of 0.9 based on a closed-form (heuristic) solution for the uniform shrinkage required for unpenalised logistic regression, assuming the c-statistic is 0.76 and overall risk is 0.68. A sample size of $n$ = 335 (about 241 events and 24 events per candidate predictor parameter) is that recommended for precise estimation of the overall risk (target 95% CI width of 0.1, assuming overall risk is 0.68); and $n$ = 75 provides a much smaller development dataset for comparison (about 51 events and 5 events per candidate predictor parameter).

### 3.2.2 Key findings and learning points

Table 1, Table S1 and Table 2 summarise the findings for development sample sizes of $n$ = 75, 355 and 456, respectively. They provide results of anticipated model performance, degradation and instability for various modelling strategies. Figure 4 and Figure 5 provide graphical displays of the posterior distributions for a few of the modelling strategies and performance metrics. We now summarise the key findings as a series of learning points. Note that, in this particular application the net benefit results for the model were very similar to those of the winner (see Figure 2), so we only discuss the former here.

***Model degradation increases as sample size decreases***

Regardless of the model development strategy, smaller sample sizes lead to CPMs with lower expected performance (higher expected model degradation), which agrees with recent research.[30] For example, consider the Bayesian lasso regression. The model's expected c-statistic is 0.69 at $n$ = 75 and 0.74 at $n$ = 465, corresponding to an expected degradation in the c-statistic of 0.07 and 0.02, respectively. In terms of net benefit, the model's ERVSI is 90% at $n$ = 75 and 97% at $n$ = 465. Lowering the sample size can also adversely impact the expected calibration slope, even when using penalisation or ensemble methods that aim to address overfitting. For example, when using a frequentist ridge regression to develop the CPM, the expected calibration slope is 1.83 at $n$ = 75 and 1.17 at $n$ = 465.

***Variability in performance and degradation increases as sample size decreases***

For all development strategies, variability in CPM performance increases as sample size decreases.

This is important because, even if the expected value is acceptable (e.g., calibration slope of 1), larger variability reduces assurance that the CPM will perform well in one particular realisation of a development dataset. For example, when using the frequentist lasso approach with $n$ = 75, the CPM has an expected calibration slope of 0.93, and thus quite close to 1. However, the posterior distribution for the calibration slope has a 95% range from -0.03 to 3.23 (Figure 5), indicating potentially large miscalibration in any one particular dataset. In contrast, with $n$ = 456, the expected value is 1.03 (thus, still close to 1) and the 95% range of 0.73 to 1.45 is much narrower (though still quite wide). The distribution of anticipated calibration curves is thus much narrow for an $n$ of 456 compared to 75 (Figure 4).

*Assurance decreases as sample size decreases*

A major consequence of the first two points is that assurance probabilities reduce as sample size decreases. For example, the tables show lower assurance probabilities for calibration slope and RVSI when $n$ is 75 compared to 456. For example, for a CPM developed using logistic regression with a uniform (heuristic) shrinkage factor, the probability that RVSI $\geq$ 90% (i.e., the proportion of the RVSI posterior distribution with values $\geq$ 90%) is 1 when $n$ is 456 but 0.55 when $n$ is 75. Similarly, the probability(0.9 $\leq$ calibration slope $\leq$ 1.1) is 0.46 when $n$ is 456 but drops to 0.11 when $n$ is 75.

*The framework embeds and extends existing sample size criteria*

Recall, previous sample size approaches provide closed-form solutions to target expected values, such as a calibration slope of 0.9 for logistic regression or a MAPE of 0.05. These targets can still be considered in our approach; for example, Table 2 shows that the expected calibration slope for a logistic regression is 0.89, close to the 0.9 value that Riley's closed-form solution anticipates for an $n$ of 456. However, the new approach enables *any* performance and degradation metric of interest to be summarised, and reveals variability and assurance probabilities, in addition to expected values, which may strongly influence the final sample size desired.

*Addressing individual-level stability and fairness may require the largest sample sizes*

Following the previous point, a larger sample size is needed to target sufficient stability of risk estimates at the individual-level and for all subgroups. For example, Figure 4 shows prediction instability plots for lasso and random forest models, which reveal large uncertainty around individual risk estimates (even spanning about 0 to 1 for some individuals), even at an $n$ of 456 where calibration curves appear quite stable. The magnitude of uncertainty may also be quite discrepant across relevant subgroups (e.g., defined by ethnicity), and larger sample sizes may be needed to address this.[8, 13]

*Table 1: Summary of posterior distributions (based on 1000 samples) of the anticipated CPM performance, degradation and instability in a large evaluation dataset, for CPMs developed with particular modelling approaches using a sample size of 75 participants (~51 events) and 10 predictor parameters. Degradation is examined relative to the performance of the reference model shown in Figure 3, which has a slope of 1, c-statistic of 0.76, and net benefit of 0.41 in the target population. ERVSI was very similar for the winning strategy.*

|  |  | *Error and uncertainty of predictions from CPM* | | *Calibration, discrimination and clinical utility of CPM* | | |
| --- | --- | --- | --- | --- | --- | --- |
| **CPM development approach** | **Sample size approach** | **MAPE:** mean (95% range) | **95% interval width:** mean (95% range) | **Calibration slope:** mean (95% range) P(0.9<slope<1.1) P(0.85<slope<1.15) | **C-statistic:** mean (95% range) mean degradation (95% range) | **Net benefit:** mean (95% range) ERVSI (95% RVSI range) P(RVSI>90%) |
| Unpenalised logistic regression (frequentist) | Fully simulation-based | 0.15 (0.09 to 0.22) | 0.66 (0.26 to 0.94) | 0.45 (0.17 to 0.83) 0.01 0.02 | 0.68 (0.60 to 0.73) -0.08 (-0.16 to -0.03) | 0.36 (0.30 to 0.39) 87.6% (74.2% to 95.7%) 0.39 |
| Unpenalised logistic regression + heuristic shrinkage | Fully simulation-based | 0.13 (0.07 to 0.20) | 0.54 (0.35 to 0.81) | 1.17 (-5.06 to 7.58) 0.11 0.18 | 0.67 (0.33 to 0.74) -0.09 (-0.43 to -0.03) | 0.37 (0.32 to 0.39) 90.1% (79.5% to 96.0%) 0.55 |
| Ridge logistic regression (frequentist) | Fully simulation-based | 0.12 (0.07 to 0.17) | 0.41 (0.37 to 0.65) | 1.83 (0.18 to 8.14) 0.13 0.18 | 0.68 (0.54 to 0.73) -0.08 (-0.22 to -0.03) | 0.37 (0.33 to 0.39) 90.3% (80.5% to 95.6%) 0.53 |
| Ridge logistic regression (Bayesian) | Fully simulation-based | 0.11 (0.06 to 0.17) | 0.48 (0.28 to 0.76) | 1.27 (0.35 to 4.02) 0.12 0.20 | 0.70 (0.62 to 0.70) -0.06 (-0.14 to -0.02) | 0.37 (0.34 to 0.39) 91.4% (83.3% to 96.9%) 0.68 |
| Lasso logistic regression (frequentist) | Fully simulation-based | 0.14 (0.08 to 0.19) | 0.46 (0.30 to 0.70) | 0.94 (-0.03 to 3.23) 0.12 0.17 | 0.62 (0.49 to 0.72) -0.13 (-0.27 to -0.03) | 0.36 (0.32 to 0.38) 88.8% (78.1% to 94.2%) 0.32 |
| Lasso logistic regression (Bayesian) | Fully simulation-based | 0.13 (0.07 to 0.19) | 0.59 (0.16 to 0.90) | 0.53 (0.25 to 0.94) 0.03 0.04 | 0.69 (0.61 to 0.74) -0.07 (-0.15 to -0.02) | 0.37 (0.32 to 0.39) 90.1% (79.0% to 96.5%) 0.57 |
| Random forest (100 trees, depth 3) | Fully simulation-based | 0.12 (0.09 to 0.16) | 0.42 (0.27 to 0.50) | 0.97 (0.60 to 1.42) 0.40 0.57 | 0.68 (0.61 to 0.71) -0.08 (-0.15 to -0.05) | 0.37 (0.34 to 0.38) 90.6% (83.2% to 94.4%) 0.69 |
| Random forest (100 trees, depth 15) | Fully simulation-based | 0.14 (0.10 to 0.18) | 0.57 (0.33 to 0.70) | 0.56 (0.33 to 0.75) 0 0 | 0.66 (0.58 to 0.70) -0.10 (-0.17 to -0.06) | 0.36 (0.32 to 0.38) 87.5% (77.7% to 93.1%) 0.29 |

*Table 2: Summary of posterior distributions (based on 1000 samples) of the anticipated CPM performance, degradation and instability in a large evaluation dataset, for CPMs developed with particular modelling approaches using a sample size of 456 participants (~310 events) and 10 predictor parameters. Degradation is examined relative to the performance of the reference model shown in Figure 3, which has a slope of 1, c-statistic of 0.76, and net benefit of 0.41 in the target population. ERVSI was very similar for the winning strategy.*

| | | *Error and uncertainty of predictions from CPM* | | *Calibration, discrimination and clinical utility of CPM* | | |
|---|---|---|---|---|---|---|
| **CPM development approach** | **Sample size approach** | **MAPE:** mean (95% range) | **95% interval width:** mean (95% range) | **Calibration slope:** mean (95% range) P(0.9<slope<1.1) P(0.85<slope<1.15) | **C-statistic:** mean (95% range) mean degradation (95% range) | **Net benefit:** mean (95% range) ERVSI (95% RVSI range) P(RVSI>90%) |
| Unpenalised logistic regression (frequentist) | Fully simulation-based | 0.051 (0.030 to 0.075) | 0.25 (0.05 to 0.50) | 0.89 (0.69 to 1.14) 0.36 0.57 | 0.75 (0.73 to 0.76) -0.01 (-0.03 to 0) | 0.40 (0.39 to 0.40) 97.8% (95.5% to 99.4%) 1.0 |
| Unpenalised logistic regression + heuristic shrinkage | Fully simulation-based | 0.050 (0.030 to 0.073) | 0.24 (0.07 to 0.42) | 0.98 (0.74 to 1.32) 0.46 0.66 | 0.74 (0.73 to 0.75) -0.02 (-0.03 to -0.01) | 0.40 (0.39 to 0.40) 97.4% (95.0% to 98.9%) 1.0 |
| Ridge logistic regression (frequentist) | Fully simulation-based | 0.078 (0.053 to 0.107) | 0.21 (0.10 to 0.38) | 1.17 (0.80 to 1.20) 0.31 0.47 | 0.72 (0.68 to 0.74) -0.04 (-0.08 to -0.02) | 0.38 (0.37 to 0.40) 94.3% (90.9% to 97.1%) 0.99 |
| Ridge logistic regression (Bayesian) | Fully simulation-based | 0.050 (0.029 to 0.073) | 0.24 (0.07 to 0.43) | 1.00 (0.74 to 1.35) 0.50 0.70 | 0.74 (0.73 to 0.75) -0.02 (-0.03 to -0.01) | 0.40 (0.39 to 0.40) 97.4% (95.0% to 99.0%) 1.0 |
| Lasso logistic regression (frequentist) | Fully simulation-based | 0.080 (0.054 to 0.114) | 0.24 (0.10 to 0.43) | 1.03 (0.73 to 1.45) 0.47 0.63 | 0.72 (0.67 to 0.74) -0.04 (-0.08 to -0.02) | 0.38 (0.37 to 0.40) 94.1% (90.1% to 97.1%) 0.98 |
| Lasso logistic regression (Bayesian) | Fully simulation-based | 0.052 (0.031 to 0.077) | 0.25 (0.06 to 0.47) | 0.89 (0.67 to 1.16) 0.39 0.60 | 0.74 (0.73 to 0.75) -0.02 (-0.03 to -0.01) | 0.40 (0.38 to 0.40) 97.3% (94.5% to 98.9%) 1.0 |
| Random forest (100 trees, depth 3) | Fully simulation-based | 0.090 (0.079 to 0.102) | 0.21 (0.14 to 0.27) | 1.63 (1.26 to 2.09) 0 0 | 0.72 (0.71 to 0.73) -0.04 (-0.05 to -0.03) | 0.38 (0.36 to 0.39) 92.8% (89.7% to 95.6%) 0.95 |
| Random forest (100 trees, depth 15) | Fully simulation-based | 0.112 (0.099 to 0.127) | 0.52 (0.20 to 0.72) | 0.62 (0.54 to 0.71) 0 0 | 0.69 (0.67 to 0.70) -0.07 (-0.09 to -0.06) | 0.37 (0.36 to 0.38) 90.7% (87.7% to 93.2%) 0.70 |

*Figure 4: Prediction and calibration instability plot, showing anticipated uncertainty of individual risk estimates and calibration curves in a large evaluation dataset (the target population), when developing a CPM using a development sample of n = 75 or n = 456 and a modelling strategy of either: logistic regression with a lasso penalty, or random forests using 100 trees with depth 3. For ease of display, only 200 calibration curves are presented.*

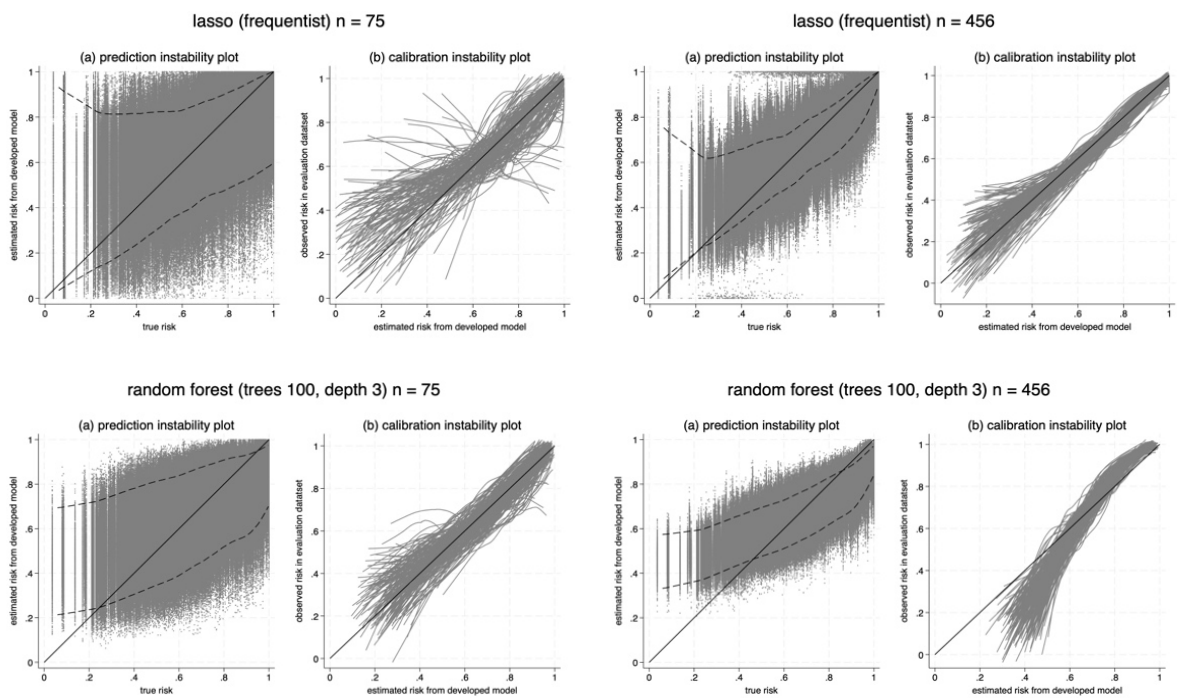

*Figure 5: Anticipated posterior distribution of calibration slope and value of information (here, the RVSI: the relative value of sample information) for the fitted model's net benefit compared to the reference model in a large evaluation dataset (target population), when using a development sample of n = 75 or n = 456 and a model development strategy of either: logistic regression with a lasso penalty, or random forests using 100 trees with depth 3*

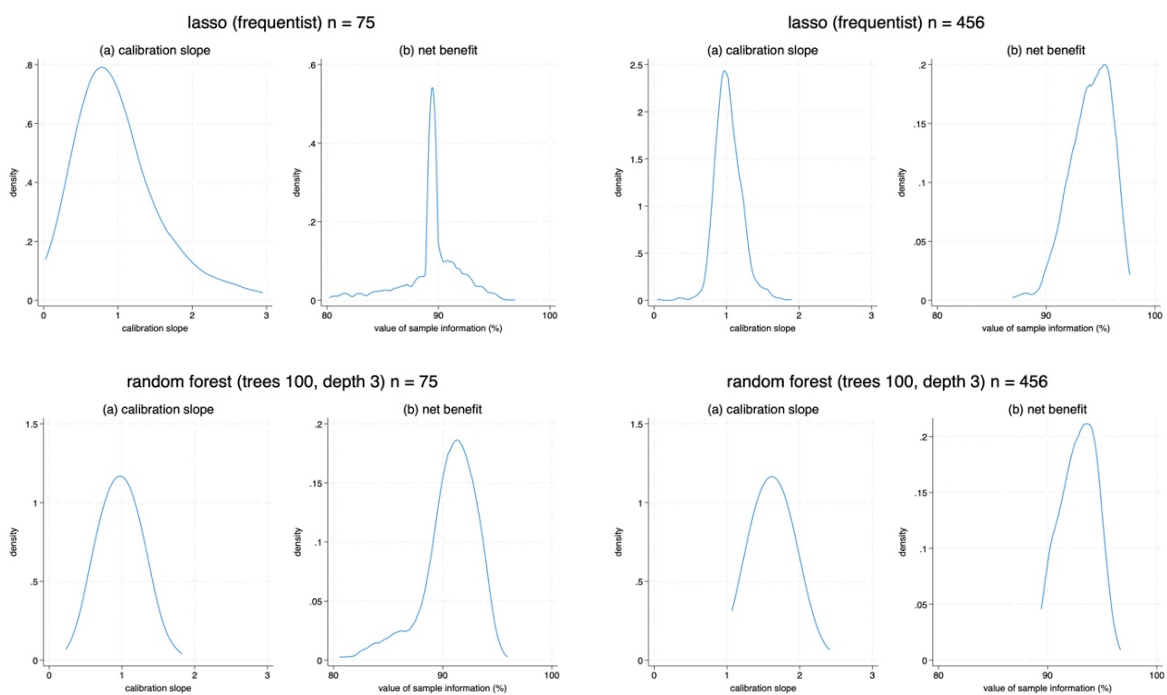

*The sample size required depends on the model development approach*

The approach used for CPM development also impacts the amount of degradation and the assurance probabilities. For example, with an $n$ of 456, the probability($0.9 \leq$ calibration slope $\leq 1.1$) is 0.70 for a Bayesian ridge regression, 0.47 for a frequentist ridge regression, and 0 for a random forest model (Table 2). Model tuning will also impact the required sample size. Tables 1 and 2 show that performance degradation and assurance probabilities are worse when using a random forest with a tree depth of 3 compared to 15. With an $n$ of 456, the expected c-statistics are 0.72 and 0.69, and the mean uncertainty interval widths for individual risk are 0.21 and 0.52, for depths of 3 and 15, respectively.

*Context and threshold choice is important*

The EVSI and ERVSI for net benefit is high, even at a small $n$ of 75, but this also reflects the threshold choice (0.5) lower than the overall prevalence (0.68). Even if a small sample leads to, say, a lasso model shrinking all predictor effects close to zero, the final model gives all individuals a risk estimate of about 0.68 plus sampling error; thus, most individuals will be above the threshold. This is akin to the fitted model becoming resembling the 'treat all' strategy, which has a net benefit of 0.36 in the evaluation dataset (88% of the reference model's net benefit); hence the fitted model's RVSI is inevitably going to be high. Had the threshold choice been, say, 0.9 then a highly penalised model would class most individuals to be below the threshold, and this more closely reflects the 'treat none' strategy, and so the model's RVSI would be closer to 0%.

*Adding candidate predictors and noise variables increase the sample size required*

The impact of a larger tree depth also illustrates a broader point: as model complexity increases, the potential for model degradation and instability increases. This also occurs when adding in additional candidate predictors, especially if many of those are noise variables. For example, Table S2 shows the impact of adding 10 extra candidate predictors, which are all considered to be noise variables, which is a conservative assumption (e.g., when researchers are exploring whether additional variables improve CPMs). At an $n$ of 75, the ERVSI for net benefit is now 0.34 and P(RVSI) $\geq$ 90% is 0.14 (Table S2), compared to 0.37 and 0.68 when the 10 noise variables were not considered (Table 1). Note that if some of the additional candidate predictors were likely to be important, then the user should then change the reference model accordingly.

*An 'acceptable' sample size depends on the chosen estimands and target/assurance values*

So what sample size is (minimally) acceptable? Given all the points above, the answer clearly depends on the user's choice of estimands, their chosen model development strategy, and the assurance guarantees they want. For example, consider a user decides to use a Bayesian ridge regression for developing the pre-eclampsia model with those 10 predictor parameters. Then, a minimum sample size of 75 might be deemed acceptable for net benefit if the user targets an RVSI of at least 80%, as there is a 0.95 probability it will be between 83% and 97% (Table 1). However, if they also target an expected MAPE of 0.05, then $n$ of 75 is insufficient (mean MAPE of 0.13, Table 1) and an $n$ of 456 is preferred (mean MAPE of 0.05, Table 2). Further, if the user wants a probability of 0.9 that $0.85 \leq$ calibration slope $\leq 1.15$, to aim for well calibrated model, then an $n$ of 456 is also insufficient (probability = 0.70) and further investigation identifies an $n$ of about 1000 is needed (probability = 0.90). Even higher sample sizes may be required to tackle fairness and stability across subgroups.[8, 13]

## 4. Improving computational time for unpenalised and penalised regression

When applying the simulation-based approach for one chosen sample size, using a 2021 MacBook Pro with 32GB ram, the time required (to produce a row of results in Tables 1 or 2) was about 45 minutes for penalised logistic regression models in a frequentist framework, and about 100 minutes in a Bayesian framework. To reduce this computational burden (and carbon footprint), in this section we use statistical theory to approximate the anticipated posterior distribution ($p(\theta_i | \boldsymbol{\beta}, \boldsymbol{x}_{new})$) for penalised regression, building on our earlier work.[8, 9]

### 4.1 Improving computational time for unpenalised logistic regression

After fitting an unpenalised logistic regression model using maximum likelihood estimation, the approximate sampling distribution for the parameter estimates ($\widehat{\boldsymbol{\beta}} = (\hat{\alpha}, \hat{\beta}_1, \hat{\beta}_2, \ldots, \hat{\beta}_P)'$) is multivariate normal. If we assume noninformative (i.e., flat improper) prior distributions, we can express the posterior distribution for $\boldsymbol{\beta}$ (the true but unknown parameters) as approximately:

$$\boldsymbol{\beta} \,|\, \mathbf{y}, \mathbf{X} \sim MVN\big(\widehat{\boldsymbol{\beta}}, \mathrm{var}(\widehat{\boldsymbol{\beta}})\big) \qquad \text{Eq. (2)}$$

where $\mathbf{X}$ is the design matrix formed using each individual's predictor values corresponding to $\boldsymbol{x_i} = (1, x_{1i}, x_{2i}, \ldots, x_{Pi})$, and $\mathbf{y}$ is the individuals' observed outcome values ($\mathbf{y}' = (y_1, y_2, \ldots, y_n)$) in the development dataset. This can be re-expressed in terms of sample size and unit information using,[8, 9]

$$\boldsymbol{\beta} \,|\, \mathbf{y}, \mathbf{X} \sim MVN\big(\widehat{\boldsymbol{\beta}}, n^{-1}\mathbf{I}^{-1}\big) \qquad \text{Eq. (3)}$$

where $\mathbf{I}$ is Fisher's *unit* information matrix ($\mathbf{I}$) and $n$ the development sample size. The unit information matrix accounts for the case-mix distribution in the development dataset and the linear predictor of the fitted model. For a logistic regression model, $\mathbf{I}$ is defined by,

$$\mathbf{I} = E\left(\frac{\exp(\mathbf{X}'\widehat{\boldsymbol{\beta}})}{\left(1 + \exp(\mathbf{X}'\widehat{\boldsymbol{\beta}})\right)^2}\mathbf{XX}'\right) \qquad \text{Eq. (4)}$$

and $E$ denotes the expected value of the matrix, and so requires averaging (i.e., obtaining mean values of) each component of the matrix across all individuals in the development dataset.

As it is independent of sample size, $\mathbf{I}$ only needs to be calculated once and can be obtained by averaging across individuals in a real (or synthetic) dataset that reflects the case-mix ($\mathbf{X}$) in the target population. This can be embedded at the end of Step (4) of the simulation-based approach, with $\mathbf{I}$ calculated based on a (large) dataset that reflects the case-mix in Step (2) and, assuming estimates are approximately unbiased, by replacing $\widehat{\boldsymbol{\beta}}$ with the $\boldsymbol{\beta}$ values of the reference model. Crucially, $\mathbf{I}$ is then fixed for subsequent sample size calculations and so the user only needs to specify $n$, their sample size of interest, to produce the anticipated posterior distribution using Eq. (3). This can be sampled from directly to obtain, say, 1000 models without requiring Steps (5) or (6) of the simulation-based approach, and allowing the application of Steps (7) to (9) to examine the predictive performance and degradation of the 1000 models when applied to the large target population dataset. The approach assumes unbiased estimates and thus is likely to perform best when the sample size is not small (we examine this later).

### 4.2 Improving computational time for penalised logistic regression

Penalised regression approaches add a penalty term to the likelihood so that predictor effect estimates are shrunk toward zero. This makes it difficult to approximate the sampling distribution of their parameter estimates in a frequentist framework; and software packages typically do not provide standard errors.

To address this, we propose a Bayesian one-sample analysis that approximates $p(\boldsymbol{\beta}|\mathbf{y},\mathbf{X})$ (the anticipated posterior distribution for $\boldsymbol{\beta}$) by combining (i) the data (likelihood) defined by anticipated parameter estimates (and the corresponding information matrix) from an unpenalised regression of a particular sample size ($n$), and (ii) prior distributions that reflect a penalised regression, for example for a lasso or ridge penalty. Specifically, we approximate,

$$p(\boldsymbol{\beta}|\mathbf{y},\mathbf{X}) \propto p(\boldsymbol{\beta})p(\mathbf{y},\mathbf{X}|\boldsymbol{\beta})$$

by replacing the likelihood of $p(\mathbf{y}|\boldsymbol{\beta},\mathbf{X})$ with $p(\widehat{\boldsymbol{\beta}}|\boldsymbol{\beta})$, where the latter is assumed $MVN(\boldsymbol{\beta}, n^{-1}\mathbf{I}^{-1})$ based on an *unpenalised* logistic regression, with $n$ a chosen sample size of interest, and $\mathbf{I}$ assumed known and obtained as described in Section 4.1. Crucially, our 'data' is now $\widehat{\boldsymbol{\beta}}$ which is simply a vector set to the $\boldsymbol{\beta}$ of the reference model. This assumes parameter estimates are unbiased, which

will be a reasonable approximation unless the development dataset is small or a relatively large number of parameters are considered. Our example confirms this below.

The choice of prior distributions ($p(\boldsymbol{\beta})$) is flexible and a broad introduction is given by van Erp et al.[31] Here, we focus on priors that reflect ridge and lasso penalties. Firstly, we assume a vague prior distribution for the intercept, $p(\alpha) \sim N(0,1000000)$. Then, we assume independent prior distributions for the predictor effects, with a penalty parameter $\lambda$ that defines the amount of shrinkage; larger values correspond to larger shrinkage. Taking a fully Bayesian approach, we also specify a prior distribution (a hyperprior) for $\lambda$ to propagate the uncertainty of $\lambda$ which can often be considerable.[32] Any priors could be chosen in principle; here, we use prior distributions that mirror ridge and lasso penalties:[33]

$$\text{Ridge: } p(\beta) \sim N(0, \lambda^2) \quad p(\lambda^2) \sim \text{inverse\_gamma}(0.01, 0.01)$$

$$\text{Lasso: } p(\beta) \sim \text{laplace}(0, \sqrt{\lambda^{-2}}) \quad p(\lambda^2) \sim \text{gamma}(1, \frac{1}{1.78})$$

We fit the Bayesian model using MCMC estimation via the Metropolis-Hastings approach, with a burn-in of 10000 and subsequently take 1000 samples (to mirror the 1000 models generated in the simulation-based approach), with thinning to reduce any autocorrelation by taking every 10$^{th}$ sample until 1000 are obtained overall. Example code is provided at https://github.com/Richard-D-Riley/code using the package *bayesmh* in Stata.

Once the posterior distribution is derived for a chosen sample size, we can sample $\boldsymbol{\beta}$ values and use them to make predictions for each individual in the large target population dataset. This leads to 1000 risk estimates for each individual (i.e., 1000 samples from their posterior distribution ($p(p_i | \boldsymbol{\beta}, x_{new})$)), which are then used in step (9) to derive posterior distributions for the model's performance and degradation in the target population.

### 4.3 Application to the pre-eclampsia example

We applied the two approaches based on Fisher's information decomposition (Sections 4.1 and 4.2) to the pre-eclampsia example. They dramatically reduce computational time, for example to about 15 minutes compared to the 100 minutes required for the fully simulation-based approach for Bayesian models. The results (shaded rows of Tables 3 and 4, S1 and S2) also suggest the approaches closely approximate the expected values of performance and degradation from the full simulation-based approach in all scenarios assessed, except for the expected value of calibration slope at $n$ of 75. Expected values of calibration slope are a much closer approximation at the larger sample sizes of $n$ of 335 or 456 (the minimum recommended by the *pmsampsize* criteria); measures of uncertainty, instability and assurance also appear a close approximation in these situations. Hence, although more detailed comparisons are needed across a wider variety of settings, the approaches provide a computationally more efficient alternative to the fully simulation-based approach for (Bayesian penalised) logistic regression models. In particular, they could be used to quickly examine multiple sample sizes, with the fuller approach used for confirmation as needed.

*Table 3: Comparison of the fully simulation-based approach and the approximation based on Fisher's information decomposition. Results summarise posterior distributions (based on 1000 samples) of the anticipated CPM performance, degradation and instability in a large evaluation dataset, for CPMs developed with particular modelling approaches using a sample size of 75 participants (~51 events) and 10 predictor parameters. Degradation is examined relative to the performance of the reference model shown in Figure 3, which has a slope of 1, c-statistic of 0.76, and net benefit of 0.41 in the target population. Each shaded row gives results for the approximation method (to improve computational speed) to the fully simulation-based approach of the previous row. ERVSI was very similar for the winning strategy.*

| | | *Error and uncertainty of predictions from CPM* | | *Calibration, discrimination and clinical utility of CPM* | | |
|---|---|---|---|---|---|---|
| **CPM development approach** | **Sample size approach** | MAPE: mean (95% range) | 95% interval width: mean (95% range) | Calibration slope: mean (95% range) P(0.9<slope<1.1) P(0.85<slope<1.15) | C-statistic: mean (95% range) mean degradation (95% range) | Net benefit: mean (95% range) ERVSI (95% RVSI range) P(RVSI>90%) |
| Unpenalised logistic regression (frequentist) | Fully simulation-based | 0.15 (0.09 to 0.22) | 0.66 (0.26 to 0.94) | 0.45 (0.17 to 0.83) 0.01 0.02 | 0.68 (0.60 to 0.73) -0.08 (-0.16 to -0.03) | 0.36 (0.30 to 0.39) 87.6% (74.2% to 95.7%) 0.39 |
| Unpenalised logistic regression (frequentist) | Frequentist approximation via Fisher's information decomposition | 0.12 (0.07 to 0.18) | 0.56 (0.29 to 0.83) | 0.57 (0.34 to 0.89) 0.02 0.03 | 0.70 (0.60 to 0.74) -0.06 (-0.16 to -0.02) | 0.36 (0.29 to 0.39) 89.2% (72.0% to 96.5%) 0.55 |
| Ridge logistic regression (Bayesian) | Fully simulation-based | 0.11 (0.06 to 0.17) | 0.48 (0.28 to 0.76) | 1.27 (0.35 to 4.02) 0.12 0.20 | 0.70 (0.62 to 0.70) -0.06 (-0.14 to -0.02) | 0.37 (0.34 to 0.39) 91.4% (83.3% to 96.9%) 0.68 |
| Ridge logistic regression (Bayesian) | Bayesian approximation via Fisher's information decomposition | 0.15 (0.08 to 0.22) | 0.44 (0.31 to 0.66) | 0.89 (-0.91 to 2.43) 0.15 0.23 | 0.61 (0.43 to 0.73) -0.15 (-0.33 to -0.03) | 0.35 (0.24 to 0.39) 86.1% (58.8% to 94.4%) 0.28 |
| Lasso logistic regression (Bayesian) | Fully simulation-based | 0.13 (0.07 to 0.19) | 0.59 (0.16 to 0.90) | 0.53 (0.25 to 0.94) 0.03 0.04 | 0.69 (0.61 to 0.74) -0.07 (-0.15 to -0.02) | 0.37 (0.32 to 0.39) 90.1% (79.0% to 96.5%) 0.57 |
| Lasso logistic regression (Bayesian) | Bayesian approximation via Fisher's information decomposition | 0.12 (0.07 to 0.19) | 0.55 (0.34 to 0.81) | 0.65 (0.31 to 1.06) 0.06 0.11 | 0.68 (0.57 to 0.74) -0.08 (-0.18 to -0.02) | 0.36 (0.29 to 0.39) 88.2% (71.5% to 96.5%) 0.48 |

*Table 4: Comparison of the fully simulation-based approach and the approximation based on Fisher's information decomposition. Results summarise posterior distributions (based on 1000 samples) of the anticipated CPM performance, degradation and instability in a large evaluation dataset, for CPMs developed with particular modelling approaches using a sample size of 456 participants (~310 events) and 10 predictor parameters. Degradation is examined relative to the performance of the reference model shown in Figure 3, which has a slope of 1, c-statistic of 0.76, and net benefit of 0.41 in the target population. Each shaded row gives results for the approximation method (to improve computational speed) to the fully simulation-based approach of the previous row. ERVSI was very similar for the winning strategy.*

| | | *Error and uncertainty of predictions from CPM* | | *Calibration, discrimination and clinical utility of CPM* | | |
|---|---|---|---|---|---|---|
| **CPM development approach** | **Sample size approach** | **MAPE:** mean (95% range) | **95% interval width:** mean (95% range) | **Calibration slope:** mean (95% range) P(0.9<slope<1.1) P(0.85<slope<1.15) | **C-statistic:** mean (95% range) mean degradation (95% range) | **Net benefit:** mean (95% range) ERVSI (95% RVSI range) P(RVSI>90%) |
| Unpenalised logistic regression (frequentist) | Fully simulation-based | 0.051 (0.030 to 0.075) | 0.25 (0.05 to 0.50) | 0.89 (0.69 to 1.14) 0.36 0.57 | 0.75 (0.73 to 0.76) -0.01 (-0.03 to 0) | 0.40 (0.39 to 0.40) 97.8% (95.5% to 99.4%) 1.0 |
| Unpenalised logistic regression (frequentist) | Frequentist approximation via Fisher's information decomposition | 0.050 (0.031 to 0.074) | 0.24 (0.06 to 0.43) | 0.91 (0.72 to 1.16) 0.41 0.65 | 0.74 (0.73 to 0.75) -0.02 (-0.03 to -0.01) | 0.40 (0.39 to 0.40) 97.5% (95.2% to 98.9%) 1.0 |
| Ridge logistic regression (Bayesian) | Fully simulation-based | 0.050 (0.029 to 0.073) | 0.24 (0.07 to 0.43) | 1.00 (0.74 to 1.35) 0.50 0.70 | 0.74 (0.73 to 0.75) -0.02 (-0.03 to -0.01) | 0.40 (0.39 to 0.40) 97.4% (95.0% to 99.0%) 1.0 |
| Ridge logistic regression (Bayesian) | Bayesian approximation via Fisher's information decomposition | 0.055 (0.030 to 0.080) | 0.24 (0.10 to 0.42) | 1.10 (0.84 to 1.48) 0.44 0.62 | 0.74 (0.72 to 0.75) -0.02 (-0.04 to -0.01) | 0.40 (0.39 to 0.40) 97.1% (94.2% to 98.9%) 1.0 |
| Lasso logistic regression (Bayesian) | Fully simulation-based | 0.052 (0.031 to 0.077) | 0.25 (0.06 to 0.47) | 0.89 (0.67 to 1.16) 0.39 0.60 | 0.74 (0.73 to 0.75) -0.02 (-0.03 to -0.01) | 0.40 (0.38 to 0.40) 97.3% (94.5% to 98.9%) 1.0 |
| Lasso logistic regression (Bayesian) | Bayesian approximation via Fisher's information decomposition | 0.051 (0.031 to 0.075) | 0.24 (0.07 to 0.44) | 0.97 (0.77 to 1.25) 0.56 0.77 | 0.74 (0.72 to 0.75) -0.02 (-0.03 to -0.01) | 0.40 (0.39 to 0.40) 97.3% (94.8% to 98.9%) 1.0 |

## 5. Accounting for uncertainty in the reference model

Inevitably, there is uncertainty about the reference model which needs to be prespecified as the basis for the calculations, for example in terms of the relative predictor weights and intercept, and the corresponding c-statistic and outcome prevalence. Therefore, researchers might want to examine the sensitivity of the calculations with other plausible reference models. This can be done by choosing a few different reference model specifications and repeating the simulation-based approach described in Section 3, to see how the required sample size changes (and perhaps then opting for the most conservative sample size identified). Alternatively, the different models could be embedded in the simulation itself. That is, the process can randomly select which reference model is used in step (3), according to specifying the selection probability for each model (e.g., models with more optimistic c-statistics might have a lower selection probability than others). More broadly, a distribution of plausible reference model parameters could be specified, to reflect uncertainty in the anticipated outcome prevalence, c-statistic and relative predictor weights. It may be pragmatic to focus on expressing uncertainty in the c-statistic (and so, keep fixed the prevalence and *relative* predictor weights).

To illustrate this, we return to the pre-eclampsia model with ten predictor parameters and consider four different reference models: the original (model 1) and three others (models 2 to 4). The relative predictor weights and true prevalence were fixed for all models (as stated for the model 1 in Section 4), but their true c-statistic (net benefit) were varied: 0.71 (0.38) for model 2, 0.76 (0.41) for model 1, 0.79 (0.42) for model 3, and 0.82 (0.44) for model 4. The probability of selecting each model was set at 0.1, 0.5, 0.3 and 0.1, respectively; this reflects a skewed distribution which gives lower chance of the extreme c-statistics and most weight to the original model, but also the potential for slightly larger c-statistics than before.

The results are shown in Table 5 for $n$ of 75 and 456, and generally similar to when assuming the original model 1 is the truth. The key differences are a slightly larger expected c-statistic (which stems from 3 of the 4 alternative reference models having larger c-statistics), and lower assurance probabilities for the net benefit, in terms of the relative value of sample information being at least 90% (e.g., probability reduced to 0.75 compared to 0.98, when $n$ is 456).

*Table 5: Summary of posterior distributions (based on 1000 samples) of the anticipated CPM performance, with and without allowing for uncertainty in the reference model (as explained in Section 5). Performance, degradation and instability are obtained from a large evaluation dataset, for CPMs developed using frequentist logistic regression with a lasso penalty and 10 predictor parameters with either 75 or 456 participants. Degradation is examined relative to the performance of the 'true model(s)'.*

| Sample size approach | Sample size | *Error and uncertainty of predictions from CPM* | | *Calibration, discrimination and clinical utility of CPM* | | |
|---|---|---|---|---|---|---|
| | | **MAPE:** mean (95% range) | **95% interval width:** mean (95% range) | **Calibration slope:** mean (95% range) P(0.9<slope<1.1) P(0.85<slope<1.15) | **C-statistic:** mean (95% range) mean degradation (95% range) | **Net benefit:** mean (95% range) ERVSI (95% RVSI range) P(RVSI>90%) |
| Fully simulation-based (one reference model) | 75 | 0.14 (0.08 to 0.19) | 0.46 (0.30 to 0.70) | 0.94 (-0.03 to 3.23) 0.12 0.17 | 0.62 (0.49 to 0.72) -0.13 (-0.27 to -0.03) | 0.36 (0.32 to 0.38) 88.8% (78.1% to 94.2%) 0.32 |
| Fully simulation-based (multiple reference models) | 75 | 0.13 (0.08 to 0.19) | 0.47 (0.31 to 0.72) | 0.87 (0 to 3.01) 0.12 0.19 | 0.64 (0.50 to 0.76) -0.13 (-0.29 to -0.04) | 0.36 (0.32 to 0.38) 87.4% (75.2% to 94.7%) 0.26 |
| Fully simulation-based (one reference model) | 456 | 0.08 (0.05 to 0.11) | 0.24 (0.10 to 0.43) | 1.03 (0.73 to 1.45) 0.47 0.63 | 0.72 (0.67 to 0.74) -0.04 (-0.08 to -0.02) | 0.38 (0.37 to 0.40) 94.1% (90.1% to 97.1%) 0.98 |
| Fully simulation-based (multiple reference models) | 456 | 0.08 (0.05 to 0.11) | 0.26 (0.12 to 0.46) | 1.01 (0.69 to 1.40) 0.48 0.63 | 0.73 (0.66 to 0.79) -0.04 (-0.09 to -0.02) | 0.38 (0.37 to 0.40) 92.3% (85.1% to 98.4%) 0.75 |

## 6. Discussion

This article proposes a general approach to sample size calculations for studies developing or updating a CPM, which embeds and generalises previous sample size proposals.[3-9] The approach can be used when deciding if an existing dataset (with a fixed sample size) is fit for purpose, or to guide the number of participants needed for recruitment to a new study. The novel focus of our article is directing researchers to anticipate the sampling or posterior distributions for model performance and degradation, conditional on a chosen sample size and model development strategy, in relation to a chosen set of candidate predictors and an assumed reference model. Further, as the fully-simulation based approach can be computationally intensive, we proposed an approximation based on statistical theory, using a decomposition of Fisher's information and a Bayesian one-sample analysis for penalised logistic regression. This substantially reduces computational burden and initial investigations demonstrate good approximation unless the effective sample size is small. It is particularly helpful for users needing to examine many different scenarios (sample sizes, modelling strategies, number of predictors etc); still a few can be confirmed by the fully-simulation based approach. Further research to improve computational speed (e.g., via parallelisation) and produce dedicated software packages is now needed.

The framework will help researchers to justify (e.g., to grant funders) their target sample size for participant recruitment, or to decide whether the sample size of an existing dataset is fit for purpose. Crucially, it allows *any* model performance metric to be examined (e.g. calibration, discrimination, clinical utility, overall fit), in terms of the expected value and variability of values anticipated in the target population; and the evaluation of model degradation by making comparisons to the reference model and determining assurance probabilities regarding likely performance and degradation in any one realised development dataset. The latter is more relevant to users than the expected value, as it is known that even when statistical and machine learning methods work well *on average*, they may have substantially lower performance in the single dataset being used for model development.[34, 35]

Clearly, our approach is most applicable when pilot or existing data are available to inform case-mix distributions, and when aiming to update or build from existing models that provide information about (relative) predictor weights, overall risk and c-statistic, so that a well-justified reference model can be specified. Without existing (pilot) data, specifying the case-mix distribution can be a challenge, though results by Pavlou et al. suggest assuming conditional independence of predictors may still be a good approximation.[7] As for any sample size calculation (e.g., for randomised trials), making assumptions about true parameter values can be difficult, but we discussed how the uncertainty of the reference model specification can be embedded in the calculation (see Section 5). Reference models have been used within other Bayesian prediction methods work,[36] and align with Sir David Cox's response to Breiman about why a transparent model (e.g., based on regression) is a helpful starting point for critical thinking.[37] Ultimately, the reference model specification is a pragmatic step toward identifying sample sizes required: once model development begins, internal validation (including prediction and classification instability checks via bootstrapping[15]) and learning curves remain important.[38]

In general, we recommend the minimum sample size recommended the Riley et al. criteria is a good starting point for these more extensive calculations,[4] as implemented in the *pmsampize* package.[28, 29] This aims to minimise overfitting in standard regression models, but the simulation-based approach allows a far broader set of criteria to be examined, including model degradation and assurance probabilities for calibration, discrimination and clinical utility. This may lead to the user identifying different required sample sizes than *pmsampsize* suggests, depending on their estimands, modelling strategies of interest, and specific target and assurance probabilities deemed acceptable. The fully simulation-based approach may also be more accurate than closed-form solutions (or the approximation based Fisher's decomposition).[7] It is important that a full set of candidate predictors (including noise variables) are reflected in the reference model and calculations, as increased model complexity will lead to greater instability, necessitating larger sample sizes (Table S2). For this reason, the reference model might include non-linear terms and interactions if they are also deemed relevant based on previous (machine learning) models in the field.

When conducting the calculations, it is important to consider the specific clinical setting and to involve key stakeholders such as patients and health professionals. This is needed to determine acceptable estimands, target values (e.g., of maximum model degradation), assurance probabilities and risk thresholds of interest. At present, discussions between model developers and stakeholders are often neglected or occur post-development. We hope this sample size approach will encourage early communication between the research team and key stakeholders, starting from the protocol writing and planning stage of the development study. We also recommend not just focusing on the overall model performance in the target population, but also the stability at the individual-level and for key subgroups, especially those identified as important for aspects of fairness (e.g., defined by protected characteristics).

In conclusion, we have proposed a general framework for examining the sample size required for prediction model development and updating, which utilises anticipated posterior (sampling) distributions conditional on a chosen sample size, set of predictors and development strategy. The approach should encourage researchers to be pro-active in examining the sample size required to target CPMs with appropriate model performance, degradation, stability, clinical utility and fairness for their clinical setting and target population of interest.

# SUPPLEMENTARY MATERIAL

*Table S1: Summary of posterior distributions (based on 1000 samples) of the anticipated CPM performance, degradation and instability in a large evaluation dataset, for CPMs developed with particular modelling approaches using a sample size of 335 participants (~241 events) and 10 predictor parameters. Degradation is examined relative to the performance of the reference model shown in Figure 3, which has a slope of 1, c-statistic of 0.76, and net benefit of 0.41 in the target population. Each shaded row gives results for an approximation (to improve computational speed) to the fully simulation-based approach of the previous row. ERVSI was very similar for the winning strategy.*

| | | *Error and uncertainty of predictions from CPM* | | *Calibration, discrimination and clinical utility of CPM* | | |
|---|---|---|---|---|---|---|
| **CPM development approach** | **Sample size approach** | **MAPE:** mean (95% range) | **95% interval width:** mean (95% range) | **Calibration slope:** mean (95% range) P(0.9<slope<1.1) P(0.85<slope<1.15) | **C-statistic:** mean (95% range) mean degradation (95% range) | **Net benefit:** mean (95% range) REVSI (95% RVSI range) P(RVSI>90%) |
| Unpenalised logistic regression (frequentist) | Fully simulation-based | 0.06 (0.036 to 0.087) | 0.29 (0.06 to 0.53) | 0.85 (0.63 to 1.14) 0.26 0.43 | 0.74 (0.72 to 0.75) -0.02 (-0.04 to -0.01) | 0.40 (0.38 to 0.40) 97.1% (94.1% to 99.3%) 1.0 |
| Unpenalised logistic regression (frequentist) | Frequentist approximation via Fisher's information decomposition | 0.059 (0.036 to 0.086) | 0.28 (0.08 to 0.50) | 0.88 (0.67 to 1.15) 0.31 0.51 | 0.74 (0.72 to 0.75) -0.020 (-0.04 to -0.01) | 0.39 (0.38 to 0.40) 96.7% (93.4% to 98.7%) 1.0 |
| Unpenalised logistic regression + heuristic shrinkage | Fully simulation-based | 0.058 (0.033 to 0.085) | 0.28 (0.09 to 0.48) | 0.98 (0.69 to 1.43) 0.40 0.56 | 0.74 (0.72 to 0.75) -0.020 (-0.04 to -0.01) | 0.39 (0.38 to 0.40) 96.7% (93.4% to 98.8%) 1.0 |
| Ridge logistic regression (frequentist) | Fully simulation-based | 0.082 (0.055 to 0.115) | 0.24 (0.12 to 0.41) | 1.19 (0.77 to 1.85) 0.27 0.41 | 0.72 (0.68 to 0.74) -0.041 (-0.08 to -0.02) | 0.38 (0.37 to 0.39) 93.9% (89.8% to 97.0%) 0.97 |
| Ridge logistic regression (Bayesian) | Fully simulation-based | 0.058 (0.035 to 0.087) | 0.27 (0.08 to 0.49) | 0.99 (0.69 to 1.39) 0.44 0.60 | 0.74 (0.72 to 0.75) -0.021 (-0.04 to -0.01) | 0.39 (0.38 to 0.40) 96.7% (93.1% to 98.7%) 1.0 |
| Ridge logistic regression (Bayesian) | Bayesian approximation via Fisher's information decomposition | 0.066 (0.036 to 0.096) | 0.26 (0.12 to 0.47) | 1.04 (0.78 to 1.47) 0.49 0.68 | 0.73 (0.70 to 0.75) -0.028 (-0.06 to -0.01) | 0.39 (0.38 to 0.40) 96.0% (92.4% to 98.6%) 1.0 |
| Lasso logistic regression (frequentist) | Fully simulation-based | 0.085 (0.057 to 0.124) | 0.27 (0.13 to 0.47) | 1.03 (0.64 to 1.50) 0.40 0.55 | 0.71 (0.66 to 0.74) -0.05 (-0.10 to -0.02) | 0.38 (0.36 to 0.39) 93.6% (88.9% to 96.8%) 0.95 |
| Lasso logistic regression (Bayesian) | Fully simulation-based | 0.061 (0.037 to 0.091) | 0.30 (0.07 to 0.54) | 0.85 (0.60 to 1.16) 0.30 0.46 | 0.74 (0.71 to 0.75) -0.022 (-0.04 to -0.01) | 0.39 (0.38 to 0.40) 96.5% (92.9% to 98.7%) 1.0 |
| Lasso logistic regression (Bayesian) | Bayesian approximation via Fisher's information decomposition | 0.059 (0.035 to 0.089) | 0.28 (0.09 to 0.50) | 0.94 (0.73 to 1.25) 0.49 0.67 | 0.74 (0.71 to 0.75) -0.022 (-0.05 to -0.01) | 0.39 (0.38 to 0.40) 96.5% (92.8% to 98.7%) 1.0 |
| Random forest (100 trees, depth 3) | Fully simulation-based | 0.092 (0.079 to 0.108) | 0.23 (0.16 to 0.30) | 1.53 (1.16 to 1.97) 0.01 0.02 | 0.72 (0.70 to 0.73) -0.04 (-0.06 to -0.03) | 0.38 (0.36 to 0.39) 93.0% (89.7% to 95.7%) 0.95 |
| Random forest (100 trees, depth 15) | Fully simulation-based | 0.115 (0.100 to 0.133) | 0.52 (0.21 to 0.71) | 0.62 (0.53 to 0.71) 0 0 | 0.69 (0.66 to 0.71) -0.07 (-0.10 to -0.06) | 0.37 (0.35 to 0.38) 90.4% (86.5% to 93.5%) 0.64 |

*Table S2: Summary of posterior distributions (based on 1000 samples) of the anticipated CPM performance, degradation and instability in a large evaluation dataset, for CPMs developed with particular modelling approaches using a sample size of 75 participants (~51 events) and 20 candidate predictor parameters (including 10 noise predictor parameters). Degradation is examined relative to the performance of the reference model shown in Figure 3, which has a slope of 1, c-statistic of 0.76, and net benefit of 0.41 in the target population. Each shaded row gives results for an approximation (to improve computational speed) to the fully simulation-based approach of the previous row. ERVSI was very similar for the winning strategy.*

| | | *Error and uncertainty of predictions from CPM* | | *Calibration, discrimination and clinical utility of CPM* | | |
|---|---|---|---|---|---|---|
| **CPM development approach** | **Sample size approach** | **MAPE:** mean (95% range) | **95% interval width:** mean (95% range) | **Calibration slope:** mean (95% range) P(0.9<slope<1.1) P(0.85<slope<1.15) | **C-statistic:** mean (95% range) mean degradation (95% range) | **Net benefit:** mean (95% range) REVSI (95% RVSI range) P(RVSI>90%) |
| Unpenalised logistic regression (frequentist) | Fully simulation-based | 0.24 (0.16 to 0.36) | 0.94 (0.56 to 1.0) | 0.17 (0 to 0.38) 0 0 | 0.63 (0.54 to 0.70) -0.12 (-0.22 to -0.06) | 0.32 (0.25 to 0.36) 78.3% (61.9% to 89.5%) 0.02 |
| Unpenalised logistic regression (frequentist) | Frequentist approximation via Fisher's information decomposition | 0.16 (0.12 to 0.22) | 0.73 (0.28 to 0.91) | 0.39 (0.19 to 0.56) 0 0 | 0.66 (0.58 to 0.72) -0.10 (-0.18 to -0.04) | 0.34 (0.28 to 0.37) 83.4% (67.6% to 92.0%) 0.10 |
| Unpenalised logistic regression + heuristic shrinkage | Fully simulation-based | 0.18 (0.11 to 0.33) | 0.84 (0.46 to 0.99) | 0.41 (-2.19 to 3.95) 0.04 0.06 | 0.62 (0.37 to 0.70) -0.13 (-0.38 to -0.06) | 0.34 (0.28 to 0.38) 83.8% (68.0% to 92.2%) 0.11 |
| Ridge logistic regression (frequentist) | Fully simulation-based | 0.13 (0.09 to 0.18) | 0.45 (0.32 to 0.63) | 2.25 (0 to 7.30) 0.13 0.20 | 0.67 (0.51 to 0.72) -0.09 (-0.25 to -0.04) | 0.37 (0.33 to 0.38) 90.1% (81.8% to 94.1%) 0.51 |
| Ridge logistic regression (Bayesian) | Fully simulation-based | 0.19 (0.11 to 0.34) | 0.86 (0.37 to 1.00) | 0.41 (0.01 to 1.21) 0.04 0.05 | 0.65 (0.56 to 0.70) -0.11 (-0.20 to -0.05) | 0.34 (0.28 to 0.37) 84.0% (69.5% to 92.0%) 0.14 |
| Ridge logistic regression (Bayesian) | Bayesian approximation via Fisher's information decomposition | 0.17 (0.12 to 0.22) | 0.46 (0.36 to 0.61) | 0.51 (-0.74 to 1.52) 0.11 0.15 | 0.57 (0.43 to 0.67) -0.19 (-0.33 to -0.08) | 0.33 (0.22 to 0.37) 82.2% (54.2% to 90.6%) 0.05 |
| Lasso logistic regression (frequentist) | Fully simulation-based | 0.14 (0.09 to 0.19) | 0.48 (0.36 to 0.68) | 0.97 (0 to 3.51) 0.11 0.16 | 0.62 (0.50 to 0.71) -0.14 (-0.26 to -0.05) | 0.36 (0.32 to 0.38) 88.8% (78.8% to 93.2%) 0.24 |
| Lasso logistic regression (Bayesian) | Fully simulation-based | 0.21 (0.12 to 0.34) | 0.88 (0.36 to 1.00) | 0.26 (0.02 to 0.57) 0 0 | 0.65 (0.57 to 0.71) -0.11 (-0.19 to -0.05) | 0.34 (0.28 to 0.37) 82.5% (69.0% to 92.1%) 0.093 |
| Lasso logistic regression (Bayesian) | Bayesian approximation via Fisher's information decomposition | 0.16 (0.11 to 0.21) | 0.69 (0.48 to 0.86) | 0.43 (0.11 to 0.72) 0 0 | 0.64 (0.53 to 0.71) -0.12 (-0.23 to -0.05) | 0.33 (0.26 to 0.38) 82.1% (64.0% to 92.3%) 0.105 |
| Random forest (100 trees, depth 3) | Fully simulation-based | 0.14 (0.11 to 0.17) | 0.38 (0.30 to 0.43) | 1.04 (0.54 to 1.51) 0.31 0.44 | 0.64 (0.57 to 0.69) -0.12 (-0.19 to -0.07) | 0.36 (0.34 to 0.38) 89.4% (83.0% to 92.3%) 0.37 |
| Random forest (100 trees, depth 15) | Fully simulation-based | 0.14 (0.11 to 0.17) | 0.47 (0.35 to 0.55) | 0.72 (0.36 to 1.01) 0.1 0.21 | 0.64 (0.56 to 0.68) -0.13 (-0.20 to -0.07) | 0.36 (0.32 to 0.37) 87.7% (78.4% to 92.0%) 0.25 |